\documentclass[conference]{IEEEtran}
\IEEEoverridecommandlockouts
\usepackage{cite}
\usepackage{amsmath,amssymb,amsfonts}
\usepackage{algorithmic}
\usepackage{graphicx}
\usepackage{textcomp}
\usepackage{paralist}
\usepackage{enumitem}
\usepackage{xcolor}
\usepackage{dblfloatfix}
\usepackage{float}
\usepackage{booktabs}
\usepackage{changes}
\def\BibTeX{{\rm B\kern-.05em{\sc i\kern-.025em b}\kern-.08em
    T\kern-.1667em\lower.7ex\hbox{E}\kern-.125emX}}
\usepackage{acronym}
\usepackage{subcaption}

\usepackage{soul}

\acrodef{5G CN}[5G CN]{5G Core Network}
\acrodef{5G-NR}[5G-NR]{5G New Radio}
\acrodef{3GPP}[3GPP]{3rd Generation Partnership Project}

\acrodef{EC}[EC]{Energy Consumption}
\acrodef{AI}[AI]{Artificial Intelligence}
\acrodef{AWS}[AWS]{Amazon Web Services}
\acrodef{ACPI}[ACPI]{Advanced Configuration and Power Interface}
\acrodef{AUSF}[AUSF]{Authentication Server Function}
\acrodef{AMF}[AMF]{Access and Mobility Management Function}
\acrodef{AI/ML}[AI/ML]{Artificial Intelligence/Machine Learning}
\acrodef{B5G}[B5G]{Beyond 5G}
\acrodef{BBU}[BBU]{Base Band Unit}

\acrodef{CNCF}[CNCF]{Cloud Native Computing Foundation}
\acrodef{CaaS}[CaaS]{Containers as a Service}
\acrodef{CI/CD}[CI/CD]{Continuous Integration/Continuous Deployment}
\acrodef{CUE}[CUE]{Carbon Usage Effectiveness}
\acrodef{CLI}[CLI]{command-line interface}
\acrodef{CUPS}[CUPS]{Control Plane and User Plane Separation}
\acrodef{CU}[CU]{Central Unit}
\acrodef{CP}[CP]{Control Plane} %

\acrodef{DC}[DC]{Data Center}
\acrodefplural{DC}{Data Centers} % Plural form
\acrodef{DU}[DU]{Distributed Unit}%s (DUs)

\acrodef{eBPF}[eBPF]{extended Berkeley Packet Filter}
\acrodef{EEGA}[EEGA]{Energy Efficient Genetic Algorithm}
\acrodef{XAI}[XAI]{eXplainable AI}

\acrodef{GA}[GA]{Genetic Algorithm}
\acrodefplural{GA}{Genetic Algorithms} % Plural form
\acrodef{gNB}[gNB]{Next-generation NodeB}
\acrodef{gNB-CU-CP}[gNB-CU-CP]{gNB-CU-Control Plane}
\acrodef{gNB-CU-UP}[gNB-CU-UP]{gNB-CU-User Plane}

\acrodef{IoT}[IoT]{Internet of Things}
\acrodef{IaaS}[IaaS]{Infrastructure as a Service}
\acrodef{ILP}[ILP]{Integer Linear Programming}
\acrodef{ITU-T}[ITU-T]{International Telecommunication Union Telecommunication Standardization Sector}
\acrodef{ICT}[ICT]{Information and Communication Technologies}
\acrodef{K8s}[K8s]{Kubernetes}
\acrodef{Kepler}[Kepler]{Kubernetes Efficient Power Level Exporter}
\acrodef{KPI}[KPI]{Key Performance Indicator}

\acrodef{LAN}[LAN]{Local Area Network}
\acrodef{LIME}[LIME]{Local Interpretable Model-agnostic Explanations}
\acrodef{LTE}[LTE]{Long Term Evolution}

\acrodef{MAC}[MAC]{Medicum Access Control}
\acrodef{ML}[ML]{Machine Learning}
\acrodef{MQTT}[MQTT]{Message Queuing Telemetry Transport}
\acrodef{MILP}[MILP]{Mixed Integer Linear Programming}
\acrodef{MNO}[MNO]{Mobile Network Operator}
\acrodef{MSE}[MSE]{Mean Squared Error}

\acrodef{NRF}[NRF]{Network Repository Function}

\acrodef{OAI}[OAI]{OpenAirInteraface}

\acrodef{PaaS}[PaaS]{Platform as a Service}
\acrodef{PUE}[PUE]{Power Utilization Effectiveness}

\acrodef{QoS}[QoS]{Quality of Service}

\acrodef{RAN}[RAN]{Radio Access Network}
\acrodef{RL}[RL]{Reinforcement Learning}
\acrodef{RLC}[RLC]{Radio Link Control}
\acrodef{RAPL}[RAPL]{Running Average Power Limit}
\acrodef{RU}[RU]{Radio Unit}
\acrodef{RF}[RF]{Random Forest}

\acrodef{SaaS}[SaaS]{Software as a Service}
\acrodef{SDG}[SDG]{Sustainable Development Goal}
\acrodef{SMF}[SMF]{Session Management Function}
\acrodef{SHAP}[SHAP]{SHapley Additive exPlanations}

\acrodef{UE}[UE]{User Equipment}
\acrodef{UN}[UN]{United Nations}
\acrodef{UP}[UP]{User Plane} %
\acrodef{UDR}[UDR]{Unified Data Repository}
\acrodef{UDM}[UDM]{Unified Data Management}
\acrodef{UPF}[UPF]{User Plane Function}
\acrodef{SDR}[SDR]{Software Defined Radio}
\acrodef{USRP}[USRP]{Universal Software Radio Peripheral}

\acrodef{VWall}[VWall]{Virtual Wall}
\acrodef{VNF}[VNF]{Virtual Network Function}
\acrodefplural{VNF}{Virtual Network Functions} % Plural
\acrodef{VM}[VM]{Virtual Machine}
\acrodefplural{VNF}{Virtual Machines} % Plural form

\acrodef{WUE}[WUE]{Water Usage Effectiveness}
\acrodef{WLAN}[WLAN]{Wireless Local Area Network}
\acrodef{WSAN}[WSAN]{Wireless Sensor and Actuator Network}

\begin{document}

%\title{PIXAR: Preliminary study on Integrating eXplainable AI for energy-efficient open RAN}
\title{Integrating Explainable AI for Energy Efficient Open Radio Access Networks}  

%\title{E3-ORAN: Explainable AI for Energy Efficiency in Open Radio Access Networks}
%\title{XENON: XAI for Energy-efficient Networks in Open RAN}

%IXAI-RAN
%
%OREO: Open RAN eXplainable AI for Energy Optimization
%Preliminary study on Integrating eXplainable AI for Energy Efficient open RAN

\setlist{noitemsep,topsep=0pt}
\vspace{-6cm}
\author{ 
\IEEEauthorblockN{L. Malakalapalli$^\$$, V. Gudepu$^\$$, B. Chirumamilla$^\dag$, S.J. Yadhunandan$^\$$ $^\bullet$, K. Kondepu$^\$$}
 \IEEEauthorblockA{$^\$$Indian Institute of Technology Dharwad, Dharwad, India. \\
 $^\dag$ Rutgers University, USA.\\
 $^\bullet$ Boston University, Boston, USA.
 }
 \IEEEauthorblockA{e-mail: cs23ms001@iitdh.ac.in, 212011003@iitdh.ac.in, k.kondepu@iitdh.ac.in}
}
\maketitle
\vspace{-0.75cm}
\begin{abstract}

The Open Radio Access Network (Open RAN) is an emerging idea ---  transforming the traditional Radio Access Networks (RAN) that are monolithic and inflexible into more flexible and innovative. 
By leveraging open standard interfaces, data collection across all RAN layers becomes feasible, paving the way for the development of energy-efficient Open RAN architectures through Artificial Intelligence / Machine Learning (AI/ML). 
However, the inherent complexity and black-box nature of AI/ML models used for energy consumption prediction pose challenges in interpreting their underlying factors and relationships. 
This work presents an integration of eXplainable AI (XAI) to understand the key RAN parameters that contribute to energy consumption. 
Furthermore, the paper delves into the analysis of RAN parameters --- \emph{airtime}, \emph{goodput}, \emph{throughput}, \emph{buffer status report}, \emph{number of resource blocks}, and many others --- identified by XAI techniques, highlighting their significance in energy consumption.
%Also, explores the XAI integration for \emph{Extended Reality (XR) use case} and scope of improving network performance through energy-based r/x Apps.
%by Kote: A statement about results if possbile
%by Venkatesh: Sure professor!
\end{abstract}

%%%%%%%Keywords%%%%%%%%


\vspace{-0.3cm}

\section{Introduction}
\vspace{-0.05cm}
The \ac{UN} aims at achieving the seventeen \acp{SDG} to create a better future for the generations to come by 2030~\cite{unsdg}. 
Among industries, \ac{ICT}, including wireless networks, plays %a crucial role in achieving all the \acp{SDG}. 
a key role in achieving SDG-13:Climate Action.
The ongoing development of \ac{B5G} networks aligns with \acp{SDG}, focusing on reducing energy consumption and carbon footprint. 
Given the \ac{ICT} sector's commitment to sustainability, there is a need to reduce energy consumption and enhance network performance.
Recent studies have highlighted that the \ac{ICT} sector contributes between 1.8\% and 2.8\% of global greenhouse gas emissions, with a significant increase in energy consumption observed in 2020~\cite{freitag2021real}.

The emergence of \ac{B5G} networks has brought about a diverse range of data-hungry applications that consume more energy. 
The \ac{RAN} faces the challenge of accommodating diverse use cases and devices while meeting stringent \ac{QoS} demands. 
The growth in data traffic, coupled with the increasing number of connected devices and resource-intensive computations are causing more energy consumption. 
\ac{RAN} itself contributes over 75\% of the total energy consumption by service provider networks~\cite{larsen2023toward}.

Energy consumption can be measured as power consumption over a period. 
To reduce energy consumption, a strategic focus on power reduction over shorter intervals emerges as a viable and effective solution.
Efficiently managing power consumption while maintaining high performance is important in light of the increasing demand for \ac{B5G} services. 
The existing literature on \ac{RAN} energy or power measurement includes various platforms, methods, and tools, all aiming to improve \ac{RAN} performance while minimizing energy consumption.

Alongside, \ac{3GPP} offers energy-saving mechanisms ---on/off strategy of base station, core network functions optimization and others --- for \acp{MNO}, however, these mechanisms need access of \ac{RAN} data for optimization.In this context,
the O-RAN Alliance's RAN Intelligent Controllers (RICs): the Non-Real-Time (Non-RT) RIC and the Near-Real-Time (Near-RT) RIC~\cite{gudepu2024drift} plays a crucial role in enabling this optimization by providing enhanced data access and control in \ac{B5G} networks. 

The Near-RT RIC supports applications to run as xApps, whereas Non-RT RIC supports rApps.
For example, rApps are used to push the policies related to the energy efficiency when to apply the energy schemes and xApps perform the corresponding actions for the required energy efficiency schemes by utilizing the AI/ML techniques.

%%%%%%%%%%%%%%%%%%%%%%%%%%%%%%%%%%%%%%%%%%%%%%%%%%%%%%%%%%%%%%%%%%%%%%%%%%
\begin{figure*}[htbp]
\centering
\includegraphics[width=0.89\linewidth]{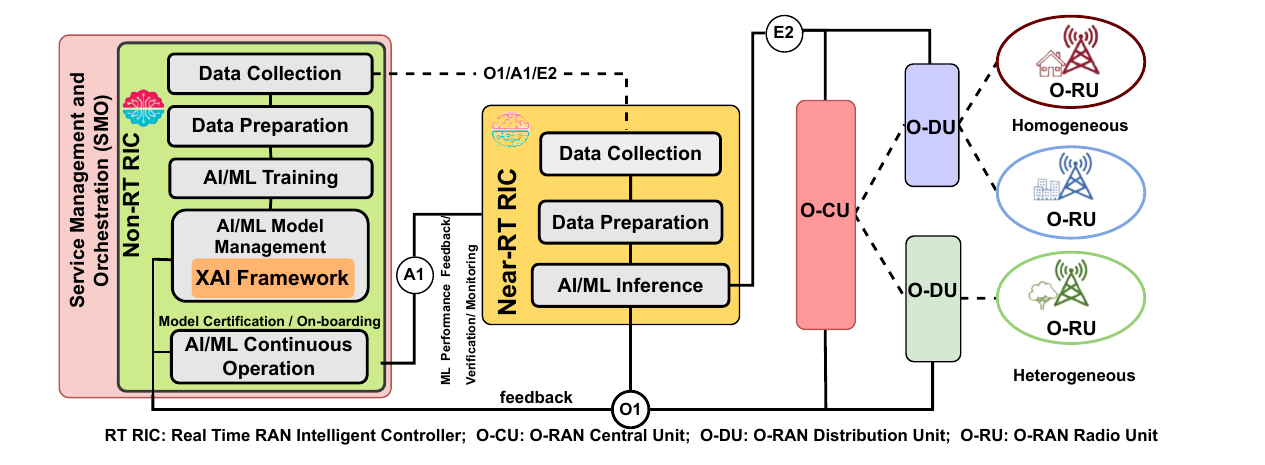}
\vspace{-0.2cm}
\caption{System Model}
\label{fig:system-model}
\vspace{-0.5cm}
\end{figure*}
%%%%%%%%%%%%%%%%%%%%%%%%%%%%%%%%%%%%%%%%%%%%%%%%%%%%%%%%%%%%%%%%%%%%%%%%%%%

The work on optimizing power consumption in the \ac{RAN} includes theoretical models, heuristic approaches, hardware and software tools, and \ac{AI/ML} based solutions~\cite{gudepu2024demonstrating}. 
% \textcolor{red}{Theoretical models and heuristics~\cite{okhovvat2023joint} aim to optimize power consumption but struggle with feasibility within reasonable time frames. 
% Hardware and software tools~\cite{gudepu2024earnest,centofanti2024impact} provide detailed power measurement and monitoring, yet face limitations such as sub-optimal solutions and limited real-time monitoring. 
% Genetic Algorithms~\cite{ilbeigi2020prediction} offer iterative improvement but are hindered by time-consuming evaluations. ~\cite{ma2022method, shapi2021energy} present the \ac{AI/ML} based power approaches for the power consumption of RAN.} 
\ac{AI/ML} approaches predict traffic patterns and network conditions, enabling dynamic resource allocation and power adjustment.
\ac{AI/ML} detects anomalies, optimizes resource allocation, and activates sleep modes during idle periods.
\ac{AI/ML} based load balancing distributes traffic efficiently, reducing power consumption while maintaining network performance.
\ac{AI/ML} driven power optimization in \ac{RAN} ensures significant energy savings and network efficiency enhancements.

However, due to the \ac{AI/ML} models complexity, black-box nature, and model bias, it is difficult to interpret \ac{AI/ML} outputs for operators when considered to oversee a decision.
The advent of \ac{XAI} techniques into \ac{RAN}, enables the transparency and interpretability for each prediction of \ac{RAN} power/energy consumption. \ac{XAI} in the realm of \ac{RAN} energy consumption offers network operators, regulators, and end-users, clear and understandable explanations for the energy-saving strategies.

This paper focuses on exploiting various \ac{RAN} parameters that contribute to \ac{RAN} energy consumption by utilizing the \ac{XAI}.
The main contributions of this paper are as follows: 
\begin{itemize}
    \item Exploring two \ac{XAI} techniques by understanding their advantages and challenges in the context of Open \ac{RAN}.
    \item Elaborating how the various \ac{RAN} parameters consume more energy by utilizing the \ac{XAI} techniques.
    \item Evaluating the considered \ac{XAI} techniques by using the real-time RAN dataset.
%    \item  The evaluation shows that the \ac{XAI} helps in understanding the energy-hungry parameters in \ac{RAN}, which requires further assistance to make the observed parameters optimized for energy.
%    \item Exploring the \emph{Extended Reality (XR)} use case in Open RAN and scope of its integration with \ac{XAI}.
\end{itemize}

%\vspace{-0.45cm}
\section{Related Work}
\vspace{-0.35cm}
\cite{okhovvat2023joint} presents the theoretical models and heuristic approaches that aim to optimize the power consumption at the \ac{RAN}.
Theoretical models --- linear models and frameworks --- face the challenge of achieving practical solutions within reasonable time frames. 
Hardware and software approaches~\cite{gudepu2024earnest} enable detailed power measurement and monitoring in various \ac{RAN} scenarios. 
The tools offer granular insights into component power usage, facilitating thorough energy pattern analysis. 
At the hardware level, tools provide detailed insights into each of the \ac{RAN} components power usage.
Various software algorithms and tools facilitate power profiling and monitoring, encompassing the measurement of energy footprints for applications and processes.
However, the sub-optimal solutions and a focus on specific power measurement levels results in limited real-time monitoring capabilities.

In~\cite{ilbeigi2020prediction}, \acp{GA} are utilized to optimize the power. 
\acp{GA} iteratively improve solutions towards optimal or near-optimal outcomes, making them useful for effective resource allocation.
Evaluating each individual in a population is time-consuming, which limits their applicability for real-time scenarios or latency sensitive use cases.

\ac{AI/ML} solutions~\cite{ma2022method, shapi2021energy} are increasingly used to optimize energy efficiency, building models for power prediction and other energy related parameters. 
The \ac{AI/ML} models excel due to their ability to predict the power consumption patterns and employ the operators to build energy-efficient networks. 
However, due to the complexity of \ac{AI/ML} models, interpreting predicted outputs poses challenges for operators. 
%Furthermore, the black-box nature of \ac{AI/ML} algorithms can make it difficult to understand the underlying factors driving the predictions, limiting the ability to troubleshoot or refine the concerned \ac{RAN} parameters.

\section{System model}
Fig.~\ref{fig:system-model} shows the O-RAN architecture that enables \emph{intelligence} through two RICs: Non- and Near-RT, operates at different components of the RAN with various time scales. 
%The Non-RT RIC oversees use cases with granularity exceeding $1\ sec$, whereas the Near-RT RIC manages those ranging between $10\ ms$ and $<\ 1 sec$.
O-RAN architecture introduces interfaces --- O1, A1, and E2 --- for facilitating data collection from the \ac{RAN} components.
O1 orchestrates components, A1 provides policy-driven guidance and AI/ML feedback, and E2 controls RAN functions via E2 control messages.

The AI/ML model management block inside the Non-RT RIC is responsible for training and deploying an AI/ML model as an x/r App.
All the collected data is stored inside the Non-RT RIC, and the AI/ML model management takes care of the AI/ML model training with the available data.
The adoption of \ac{XAI} could help to interpret the predictions of r/x Apps and RICs can optimize the energy at RAN through E2 control message by fine tuning the parameters that consume energy. 

This work investigates two popular \ac{XAI} techniques --- \ac{SHAP} and \ac{LIME} --- and their integration into the Open RAN.
\ac{SHAP}~\cite{shapi2021energy} and \ac{LIME}~\cite{dieber2020model} are techniques utilized for explaining the interpretability of \ac{AI/ML} models, known for their model-agnostic nature --- can be applied to any type of model. 
Both methods assess the impact of various parameters on model predictions.

\ac{SHAP}~\cite{shapi2021energy} values assess the significance of a feature by contrasting model predictions with and without the considered feature. 
However, because the sequence in which a model encounters features can influence its predictions, this comparison is conducted in every conceivable order to ensure a fair comparison of features.
Consequently, our approach involves evaluating all potential combinations of feature values, both with and without a particular feature, to precisely compute the \ac{SHAP} value.

\ac{LIME}~\cite{dieber2020model} calculates values through local sub-sampling of the dataset, whereas \ac{SHAP} calculates values by removing specific features and evaluating their importance.
\ac{LIME} approximates the mapping function $f(x)$ of the ML model by sampling instances, referred to as input perturbation. 
It generates synthetic samples $x_0$ closely resembling the original instance $x$, passes them to the original model $f$, and records the predictions. 
The perturbation process helps \ac{LIME} to understand how different input fluctuations affect the model output.
Ultimately, \ac{LIME} can explain a particular prediction by identifying which features contribute most significantly to it.

Regarding interpretability scope, \ac{LIME} offers localized insights ideal for simpler models.
\ac{SHAP} provides both global and local interpretability.
Both \ac{LIME} and \ac{SHAP} are model-agnostic and adaptable to various \ac{AI/ML} models.
\ac{LIME} generates local approximations through perturbing input data, while \ac{SHAP} computes Shapley values, prioritizing model-specific characteristics.
\ac{LIME} is better suited for simpler models, while \ac{SHAP} handles tasks with varying complexity due to its comprehensive understanding of feature contributions.
Regarding the stability, \ac{LIME} may be unstable due to random sampling, while the \ac{SHAP} tends to be more stable and consistent, ensuring reliability across multiple runs.
The integration of XAI for Open RAN is evaluated for a real-time dataset and obtained results are discussed in the following section.

\vspace{-0.15cm}
\section{experimental results}

A real-time dataset~\cite{Salvat2022data} from the O-RAN testbed is considered to evaluate the \ac{XAI} techniques and understand the key parameters of the \ac{RAN} that contributes energy consumption.

The dataset comprises two sets of energy measurement data: \emph{dataset\_ul} and \emph{dataset\_dlul}. 
The former captures performance and power consumption metrics of a next-generation evolved NodeB (gNB) solely utilizing the uplink channel, whereas the later includes both uplink and downlink channels. 
Configurations within each dataset detail essential parameters --- timestamp, CPU platform, \ac{LTE} interface bandwidth, transmission mode, and traffic load, among others. 
Additionally, measurements includes a comprehensive range of indicators including Modulation and Coding Scheme (MCS), Block Error Rate (BLER), throughput, power consumption, Signal-to-Noise Ratio (SNR), and clock speed. 
Both \emph{dataset\_ul} and \emph{dataset\_dlul} offers a valuable insights into the complexities of energy consumption and performance in virtualized base station deployments, enhancing the efficiency and reliability of 5G networks.

\vspace{-0.15cm}
\subsection{Analysis and Comparison of the considered \ac{AI/ML} models}
The AI/ML models that are considered are \emph{Gradient boosting}, \emph{Random Forest (RF)}, and \emph{eXtreme Gradient Boosting (XGBoost)} due to their robust performance in predictive analytics and their ability to handle complex, high-dimensional data effectively.

%%%%%%%%%%%%%%%%%%%%%%%%%%%%%%%%%%%%%%%%%%%%%%%%%%%%%%%%%%%%%%%%%%%%%%%%%%%
\begin{figure*}[htb]
    \centering
    \begin{subfigure}{0.495\linewidth}
        \centering
        \includegraphics[width=\linewidth]{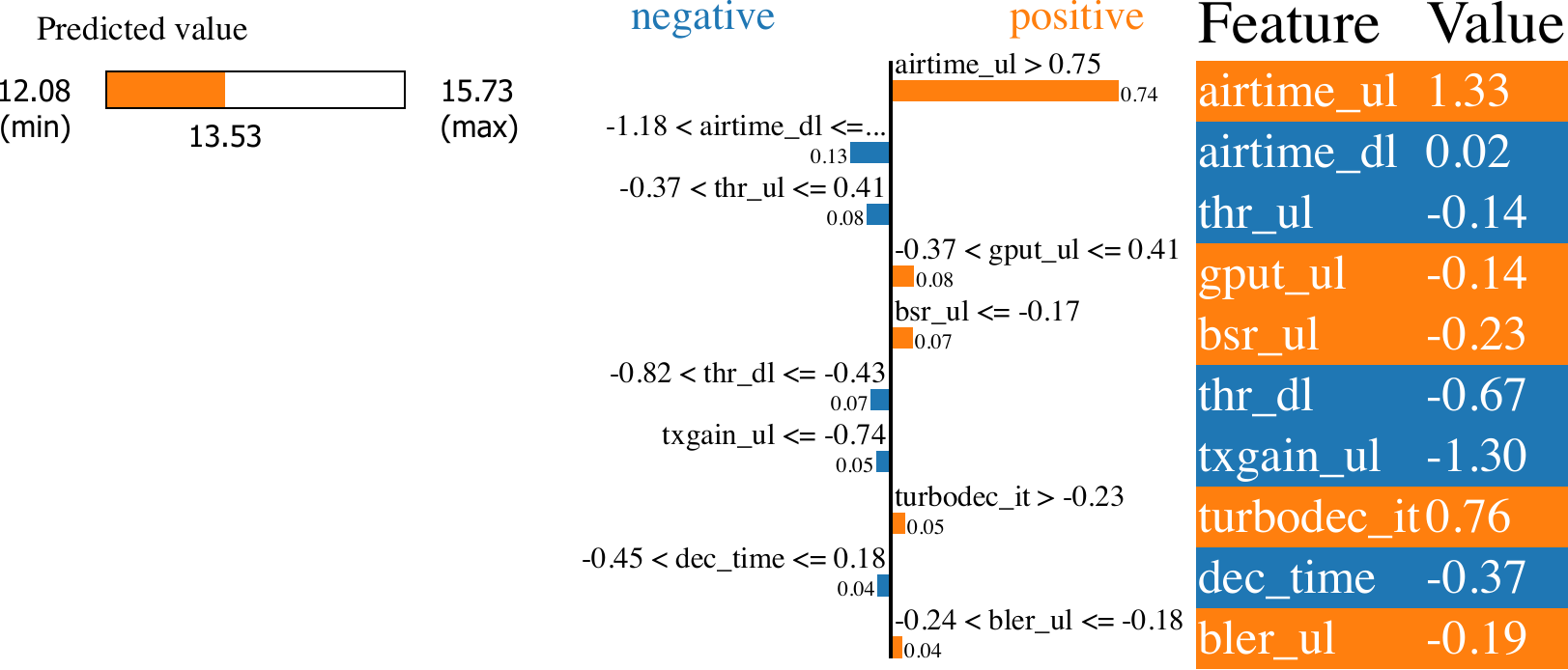}
        \caption{LIME: DL/UL dataset using Gradient Boosting}
        \label{fig:lime1}
    \end{subfigure}
    \begin{subfigure}{0.495\linewidth}
        \centering
        \includegraphics[width=\linewidth]{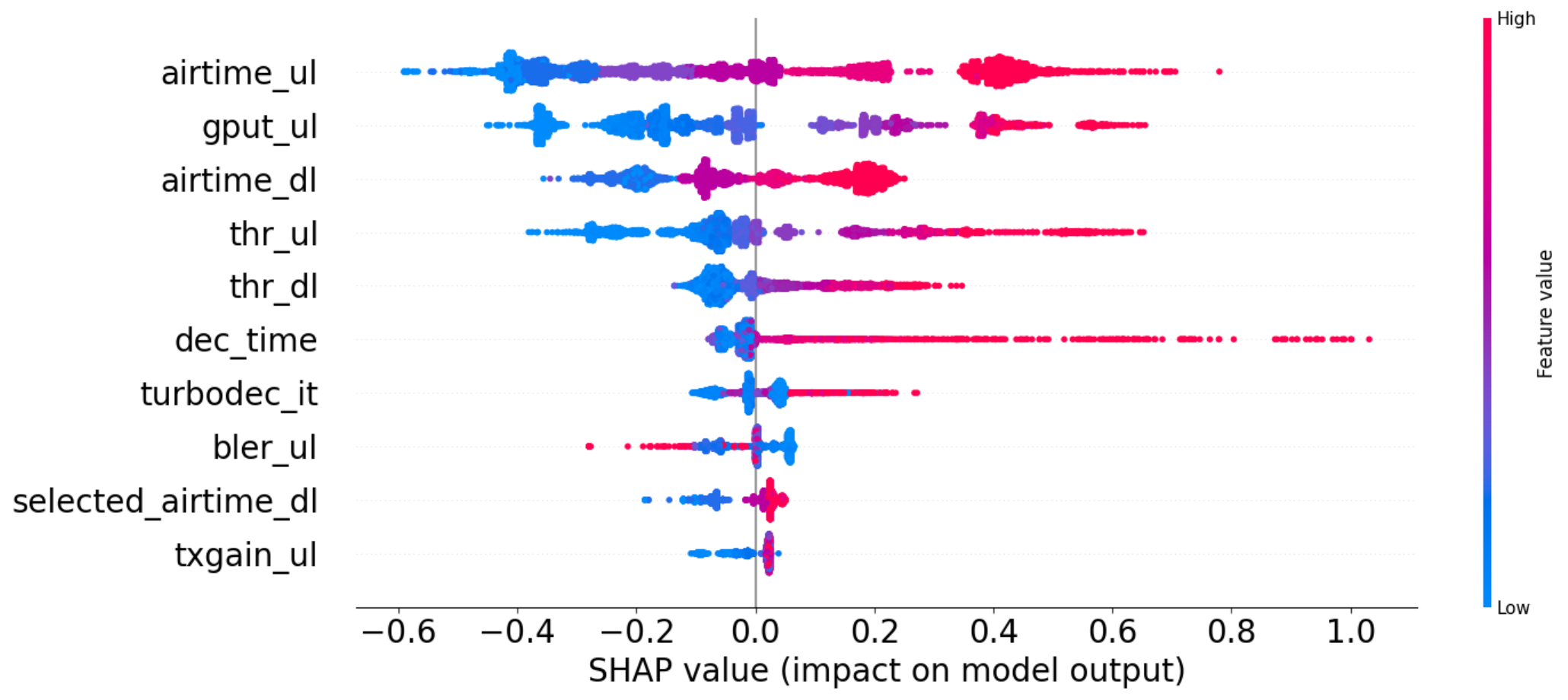}
        \caption{SHAP: DL/UL dataset using Gradient Boosting}
        \label{fig:shap1}
    \end{subfigure}
    \qquad
    \begin{subfigure}{0.495\linewidth}
        \centering
        \includegraphics[width=\linewidth]{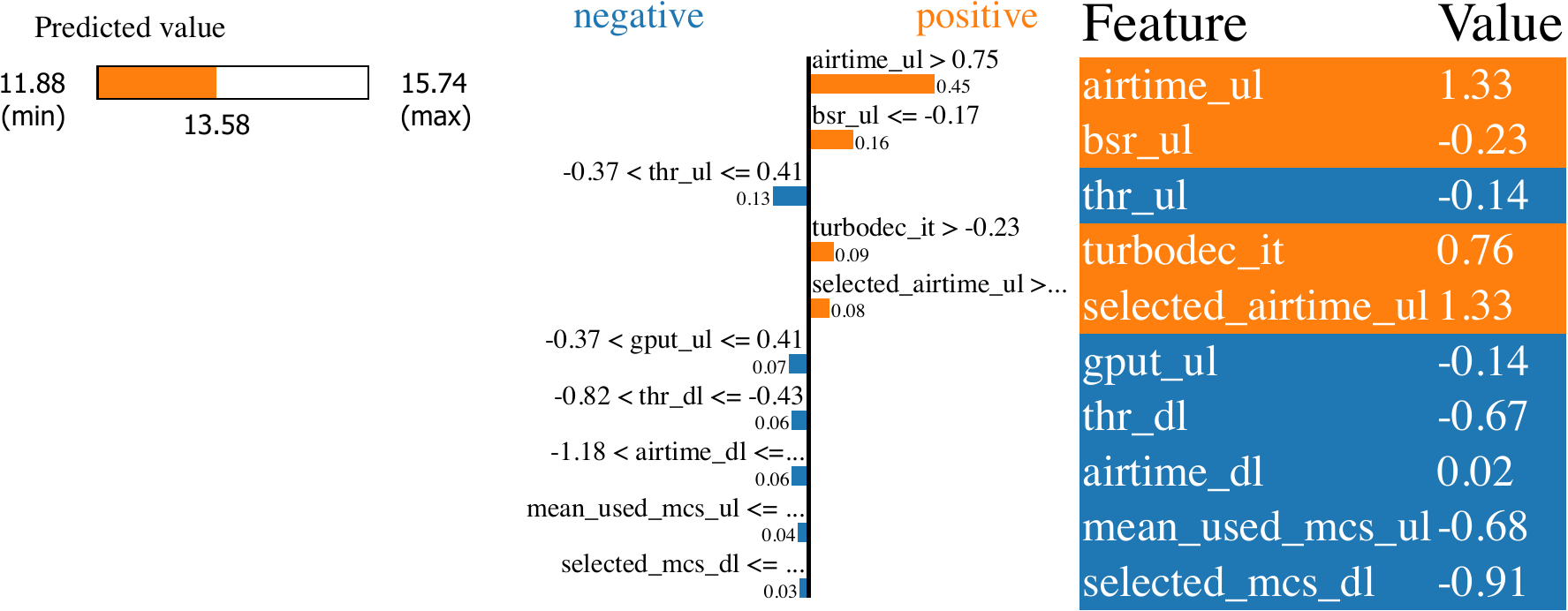}
        \caption{LIME: DL/UL dataset using Random Forest}
        \label{fig:lime3}
    \end{subfigure}
    \begin{subfigure}{0.495\linewidth}
        \centering
        \includegraphics[width=\linewidth]{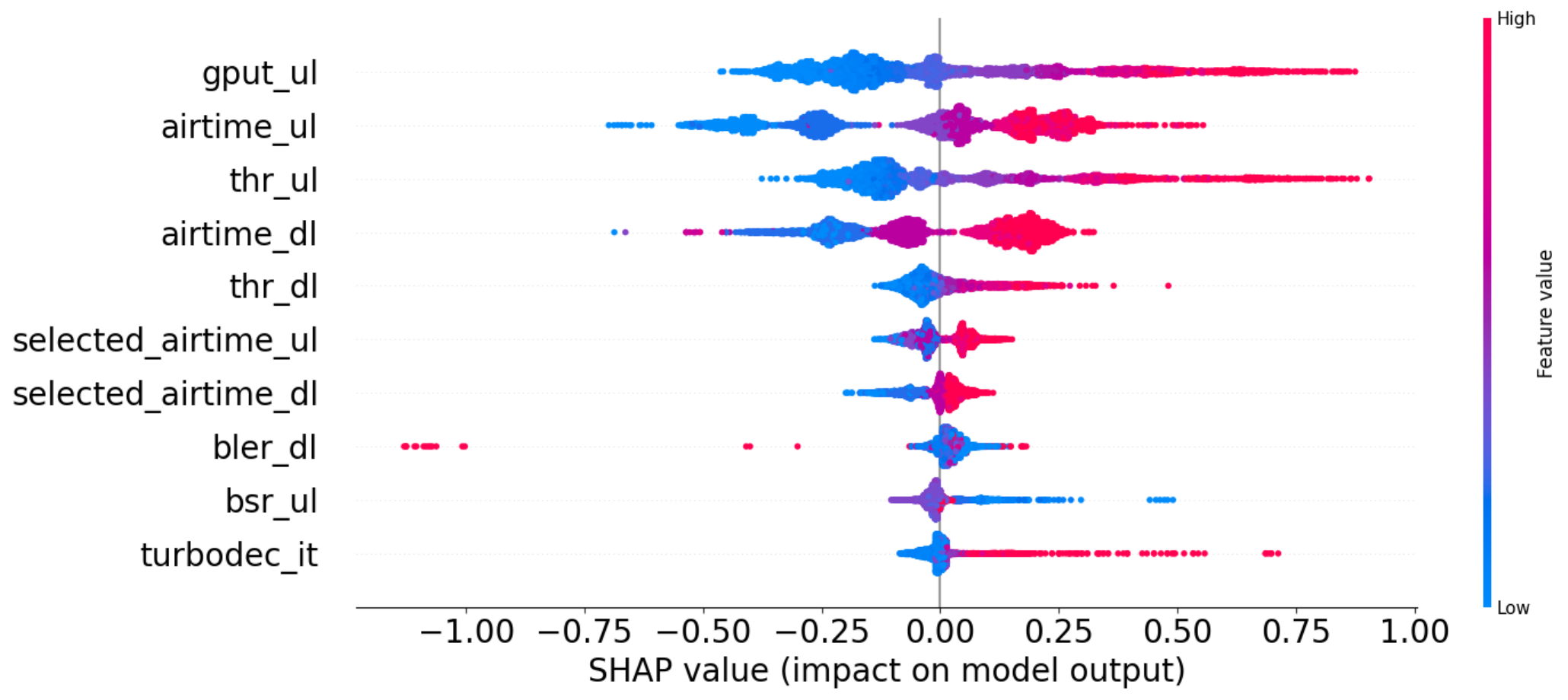}
        \caption{SHAP: DL/UL dataset using Random Forest}
        \label{fig:shap3}
    \end{subfigure}
    \qquad
    \begin{subfigure}{0.495\linewidth}
        \centering
        \includegraphics[width=\linewidth]{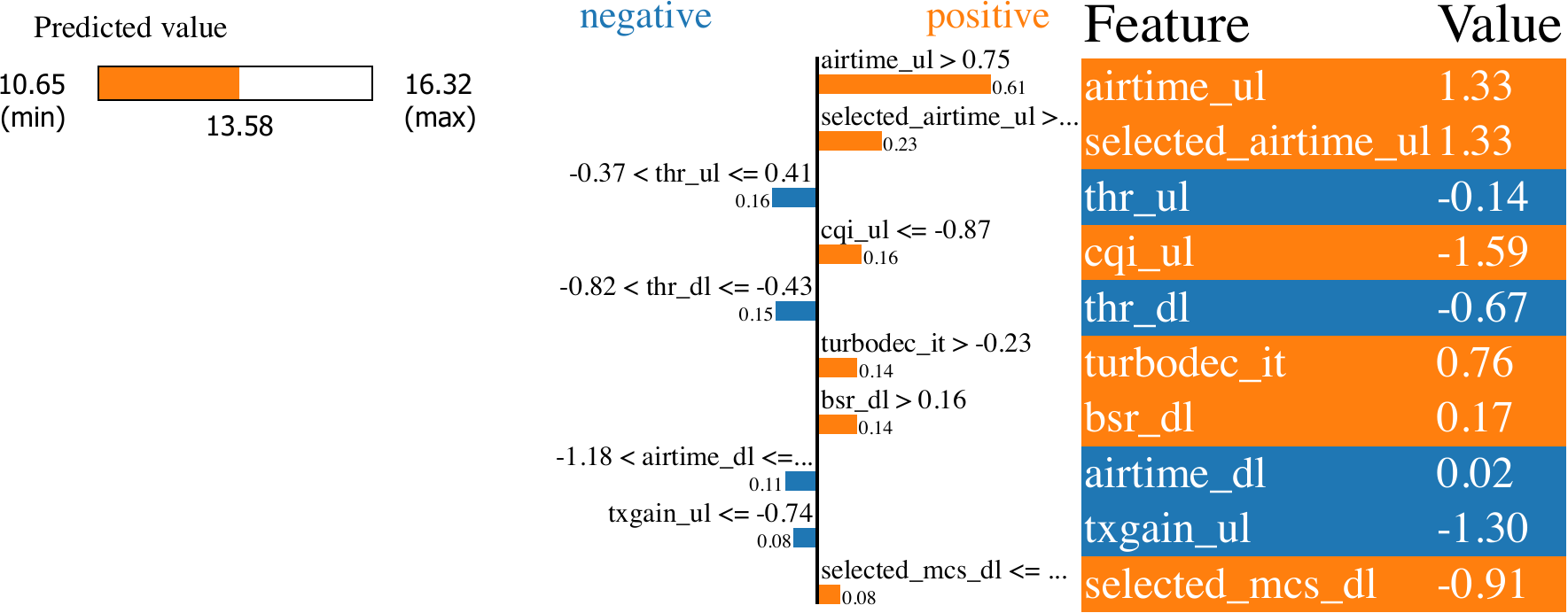}
        \caption{LIME: DL/UL dataset using XGBoost}
        \label{fig:lime5}
    \end{subfigure}
    \begin{subfigure}{0.495\linewidth}
        \centering
         \includegraphics[width=\linewidth]{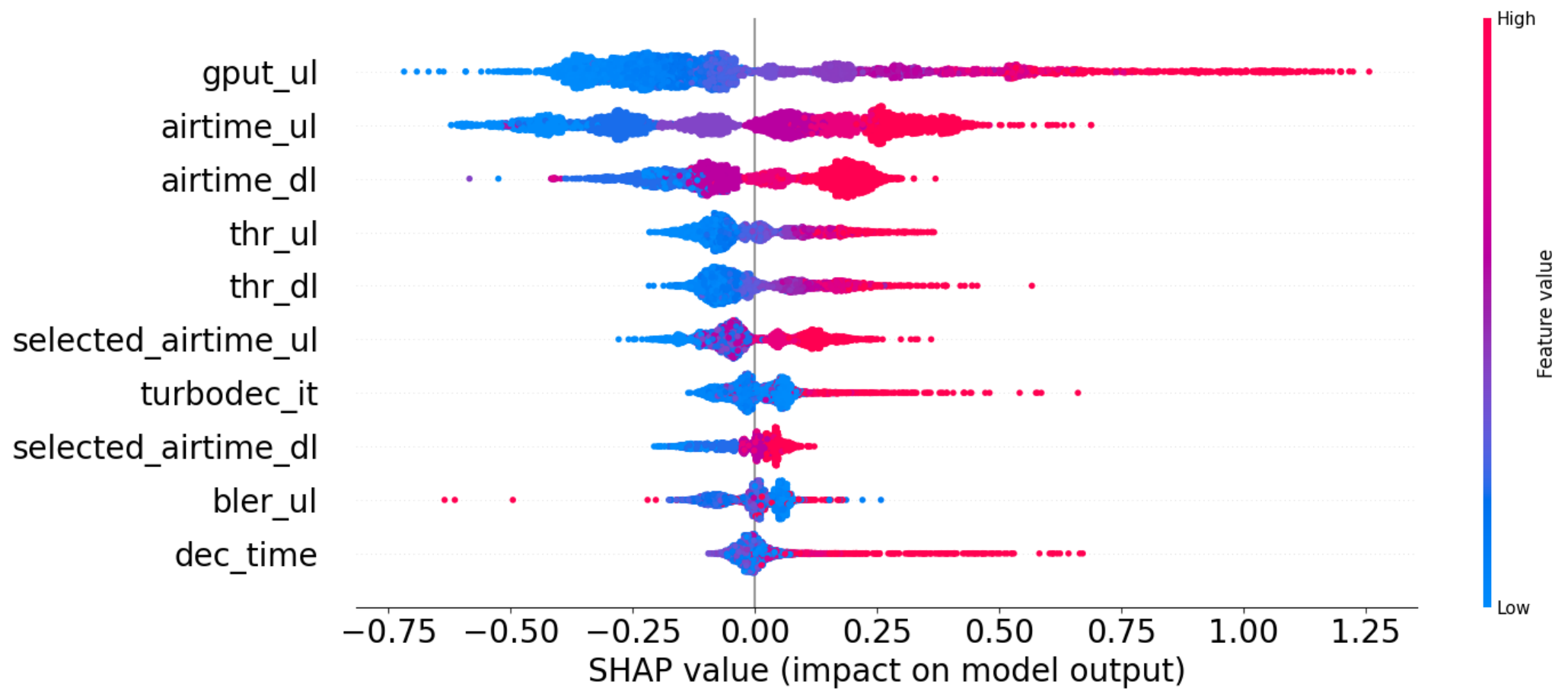}
        \caption{SHAP: DL/UL dataset using XGBoost}
        \label{fig:shap5}
    \end{subfigure}
    \qquad
   \vspace{-0.4cm}
    \caption{LIME and SHAP analysis for DL/UL RAN parameters influence on power consumption}
    \label{fig:dlul}
\vspace{-0.4cm}
\end{figure*}
%%%%%%%%%%%%%%%%%%%%%%%%%%%%%%%%%%%%%%%%%%%%%%%%%%%%%%%%%%%%%%%%%%%%%%%%%%%

\emph{Gradient boosting}~\cite{friedman2002stochastic} is a sequential \ac{AI/ML} technique that constructs predictive models iteratively to minimize prediction error. 
\emph{Gradient boosting} sets target outcomes for each model iteration based on how changes in predictions affect overall error. 
By adjusting predictions for individual cases, \emph{Gradient boosting} iteratively improves model performance, aiming to minimize error for each training case.

\emph{\ac{RF}}~\cite{breiman2001random} is an ensemble learning method comprising decision trees%~\cite{quinlan1986induction}
, constructed using random samples drawn with replacement from the training data. 
By employing bagging and feature randomness, \emph{RF} creates a set of uncorrelated decision trees. 
% Each tree involves feature selection at internal nodes, dividing the data into subsets with similar responses. 
% The best split at each node is determined by considering all input features or a random subset. 
% Features are selected based on criteria like Gini impurity or information gain\cite{decision_tree_learning}, prioritizing those leading to the greatest decrease in impurity. 
% Feature importance is assessed by the average impurity decrease across the forest. 
% Randomness aims to reduce estimator variance, curbing overfitting seen in individual trees. 
% By aggregating diverse trees, \ac{RF} decreases variance, usually enhancing model performance with a slight increase in bias.

\emph{eXtreme Gradient Boosting (XGBoost)}~\cite{chen2016xgboost} is an efficient, flexible, and scalable \ac{AI/ML} technique that extends gradient boosting.
XGBoost follows the principles of traditional gradient boosting, sequentially constructing a predictive model by aggregating the predictions of weak learners, often decision trees.
XGBoost employs a regularized learning objective function, combining a loss function to measure prediction errors with regularization terms to manage model complexity, thus preventing overfitting and enhancing generalization performance.
While traditional gradient boosting methods use first-order optimization techniques, XGBoost can optionally utilize second-order optimization techniques, such as Newton`s method, to further improve convergence speed and model accuracy. 
%XGBoost includes built-in capabilities for tree pruning and column subsampling during tree construction. 
%This helps prevent overfitting by limiting the complexity of individual trees and introducing randomness into the model. 
%XGBoost has built-in support for handling missing values in the dataset, allowing it to learn the best imputation strategy during training automatically. 

The \ac{MSE} is considered as model metrics for \emph{Gradient Boosting}, \emph{Random Forest}, and \emph{XGBoost}. 
The train \ac{MSE} and test \ac{MSE} of each model are as listed in Table~\ref{tab:model_performance_dlul} and \ref{tab:model_performance_ul} for both DL/UL and UL datasets.
Here, train \ac{MSE} indicates how well a model fits the training data and the test \ac{MSE} shows how well a model performs on unseen data. The proportions of training and test datasets are 80\% and 20\% respectively.
% As shown Table~\ref{tab:model_performance_dlul} and \ref{tab:model_performance_ul}, \emph{Gradient Boosting} has a reasonable balance between train and test \ac{MSE}, indicating good performance and generalization.
% However, \emph{Random Forest} model shows better performance on training data but overfits, as evidenced by the high test MSE.
% \emph{XGBoost} model performs well but does not fit the training data as closely as \emph{Random Forest}, and its performance on test data is similar to that of \emph{Gradient Boosting}.

%%%%%%%%%%%%%%%%%%%%%%%%%%%%%%%%%%%%%%%%%%%%%%%%%%%%%%%%%%%%%%%%%%%%%%%%%%%
\begin{table}[!htb]
\centering
\vspace{-0.3cm}
\caption{Model Performance Metrics for DL/UL dataset}
\begin{tabular}{lcc}
\toprule
\textbf{Model} & \textbf{Train \ac{MSE} [W]} & \textbf{Test \ac{MSE} [W]} \\
\midrule
Gradient Boosting & 0.05733 & 0.06710 \\
Random Forest & 0.00897 & 0.06806 \\
XGBoost & 0.01672 & 0.07021 \\
\bottomrule
\end{tabular}
\label{tab:model_performance_dlul}
\end{table}
%%%%%%%%%%%%%%%%%%%%%%%%%%%%%%%%%%%%%%%%%%%%%%%%%%%%%%%%%%%%%%%%%%%%%%%%%%%

%%%%%%%%%%%%%%%%%%%%%%%%%%%%%%%%%%%%%%%%%%%%%%%%%%%%%%%%%%%%%%%%%%%%%%%%%%%
\begin{table}[!htb]
\centering
\vspace{-0.5cm}
\caption{Model Performance Metrics for UL dataset}
\begin{tabular}{lcc}
\toprule
\textbf{Model} & \textbf{Train \ac{MSE} [W]} & \textbf{Test \ac{MSE} [W]} \\
\midrule
Gradient Boosting & 0.0290 & 0.0307 \\
Random Forest & 0.0020 & 0.0143 \\
XGBoost & 0.0085 & 0.0191 \\
\bottomrule
\end{tabular}
\label{tab:model_performance_ul}
\end{table}
%%%%%%%%%%%%%%%%%%%%%%%%%%%%%%%%%%%%%%%%%%%%%%%%%

%\vspace{-0.4cm}
\vspace{-0.15cm}
\subsection{The influence of \ac{RAN} parameters on Energy Consumption}
\begin{figure*}[htb]
    \centering
    \begin{subfigure}{0.4965\linewidth}
        \centering
         \includegraphics[width=\linewidth]{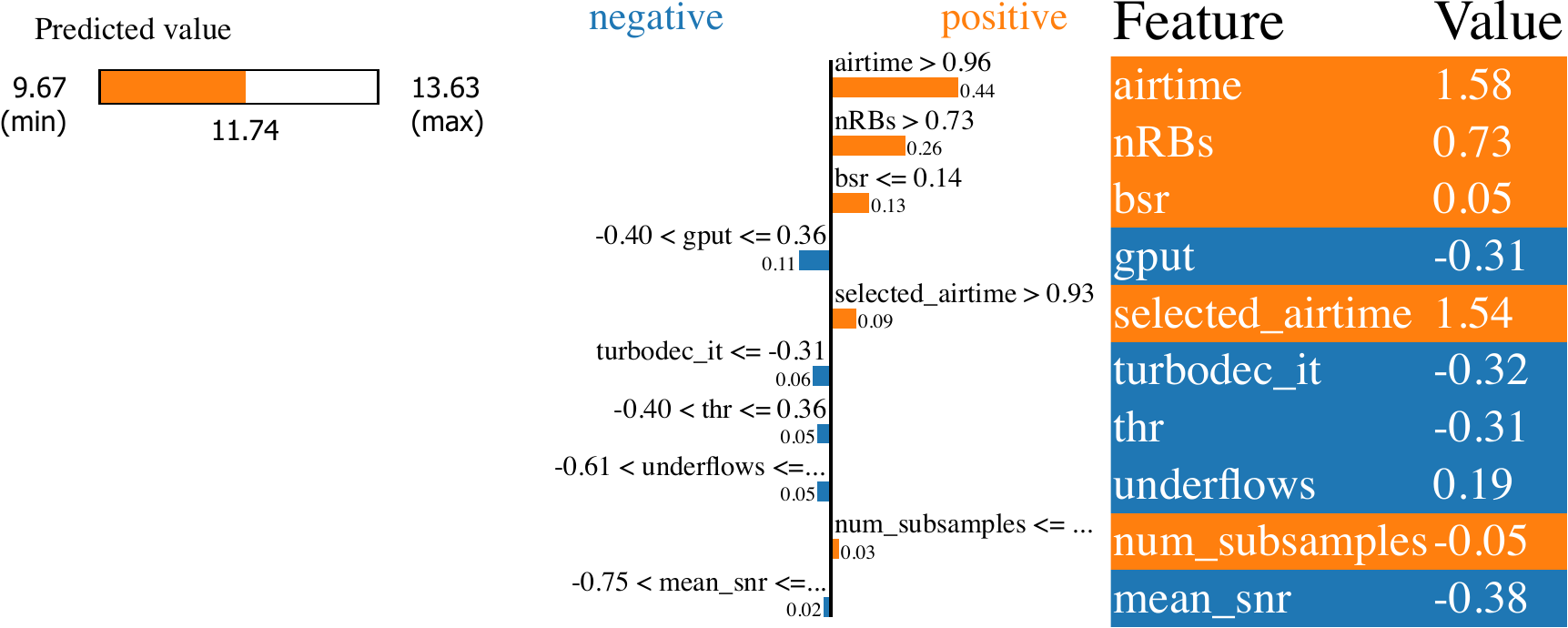}
         \caption{LIME: UL dataset using Gradient Boosting}
        \label{fig:lime2}
    \end{subfigure}
    \begin{subfigure}{0.4965\linewidth}
        \centering
        \includegraphics[width=\linewidth]{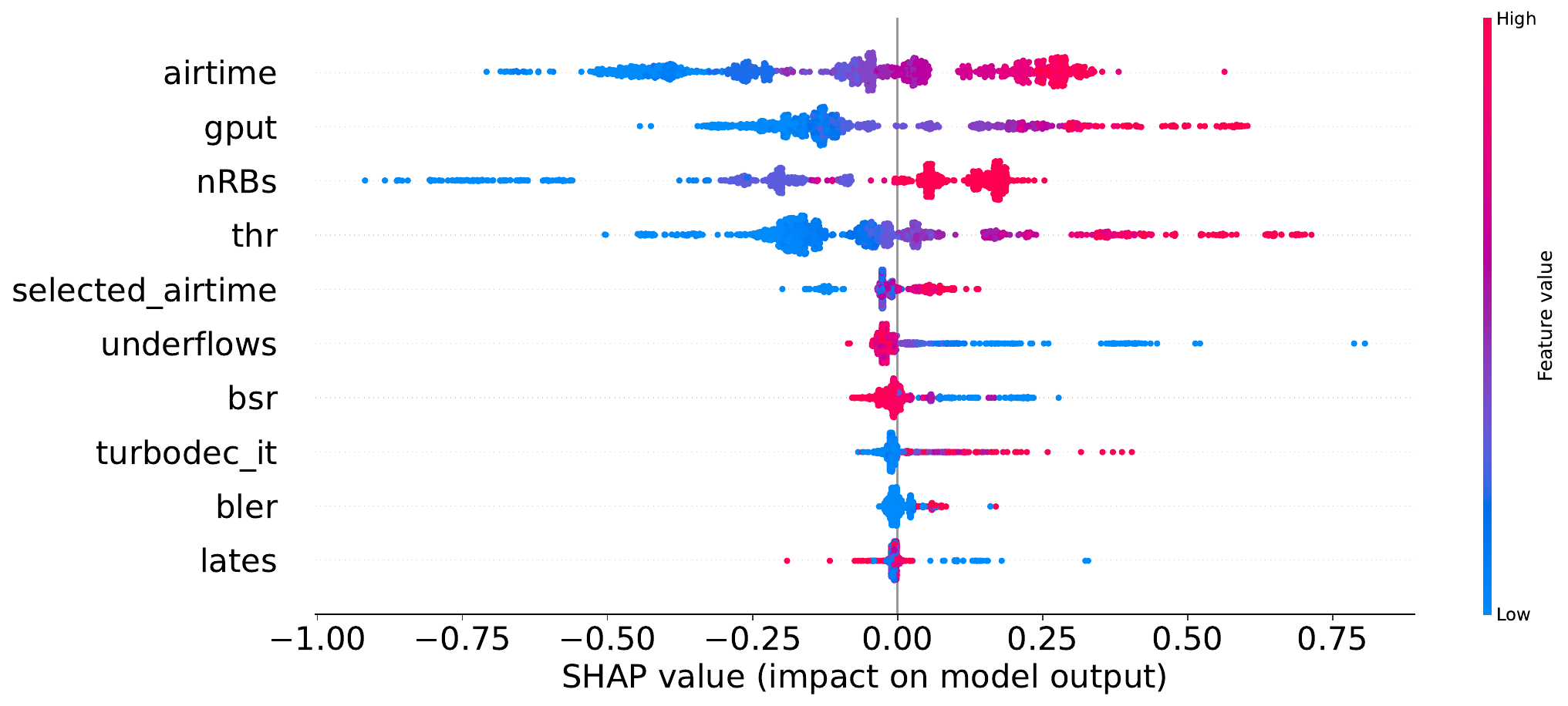}
        \caption{SHAP: UL dataset using Gradient Boosting}
        \label{fig:shap2}
    \end{subfigure}
    \qquad
    \begin{subfigure}{0.4965\linewidth}
        \centering
       \includegraphics[width=\linewidth]{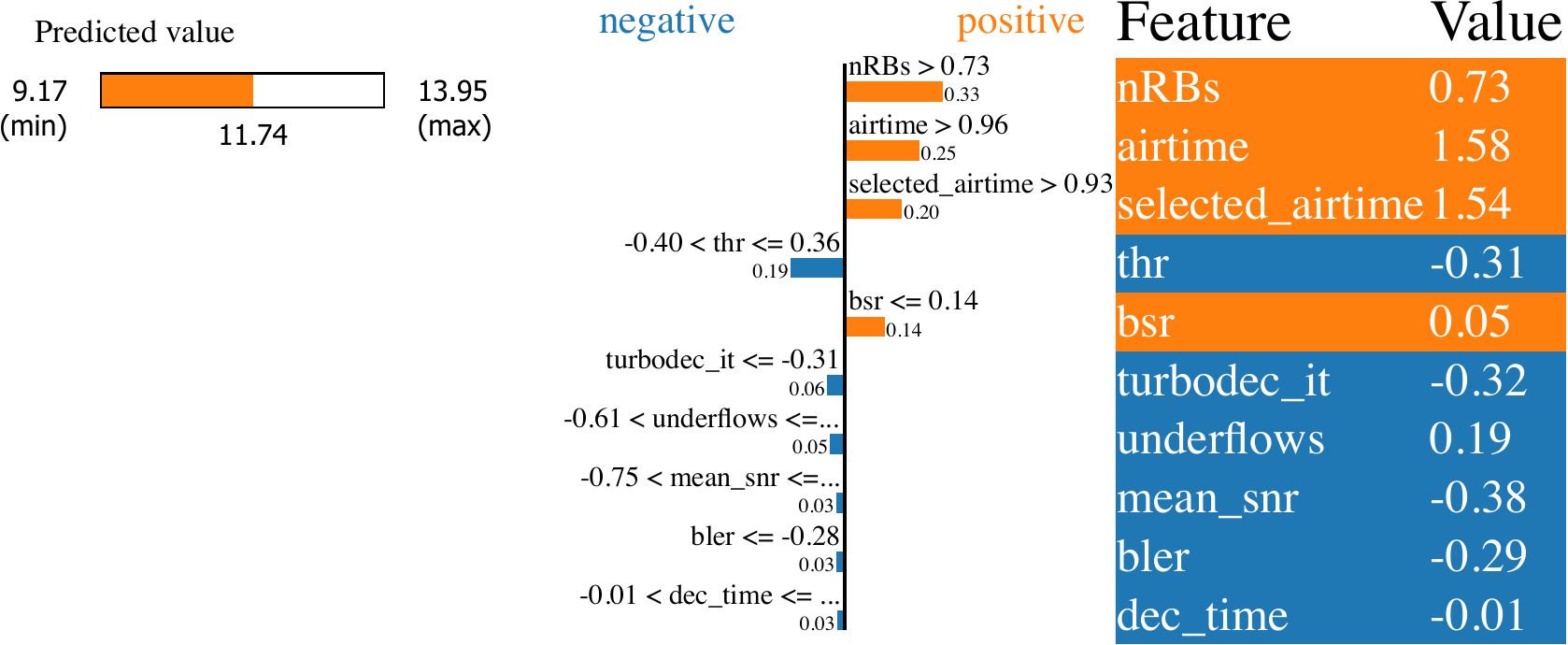}
       \caption{LIME: UL dataset using Random Forest}
        \label{fig:lime4}
    \end{subfigure}
    \begin{subfigure}{0.4965\linewidth}
        \centering
        \includegraphics[width=\linewidth]{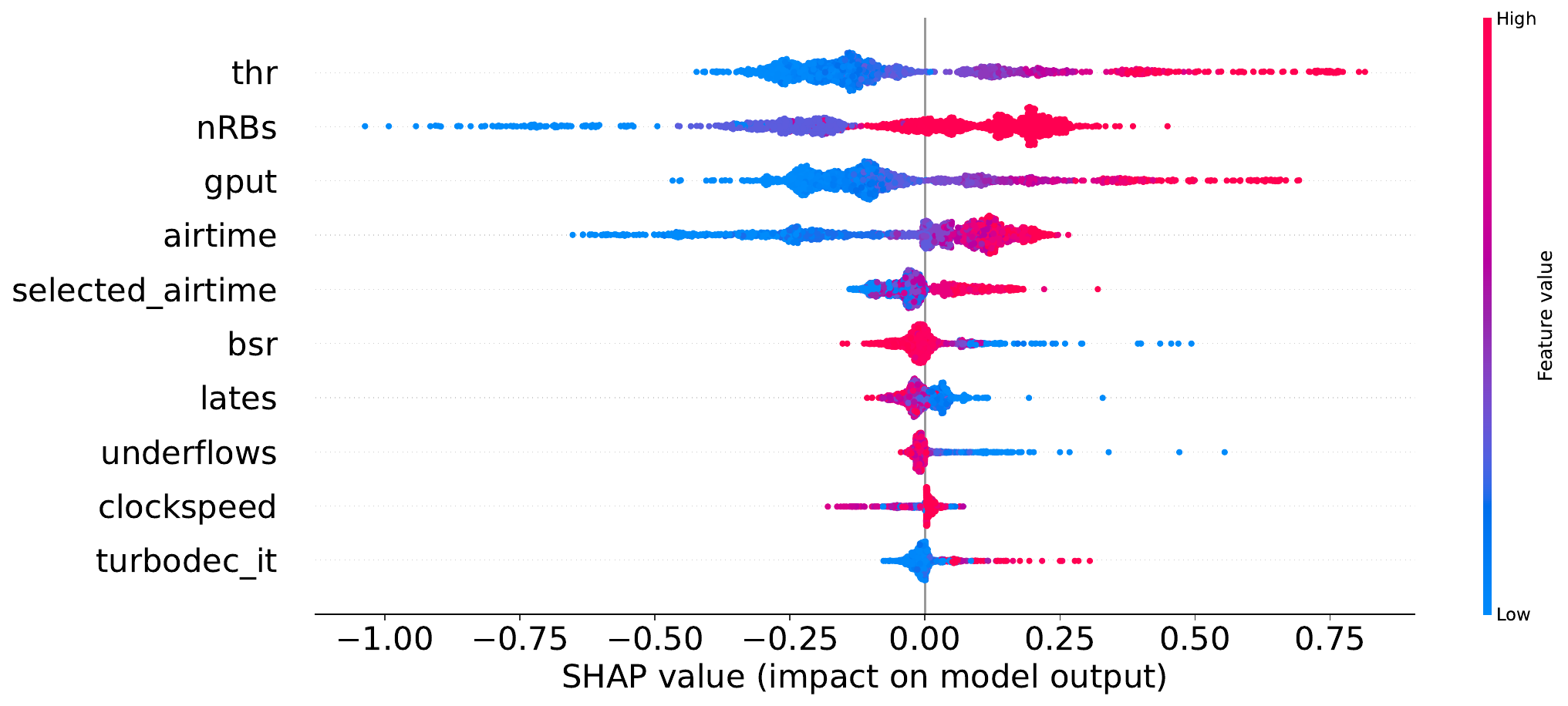}
        \caption{SHAP: UL dataset using Random Forest}
        \label{fig:shap4}
    \end{subfigure}
    \qquad
    \begin{subfigure}{0.4965\linewidth}
        \centering
       \includegraphics[width=\linewidth]{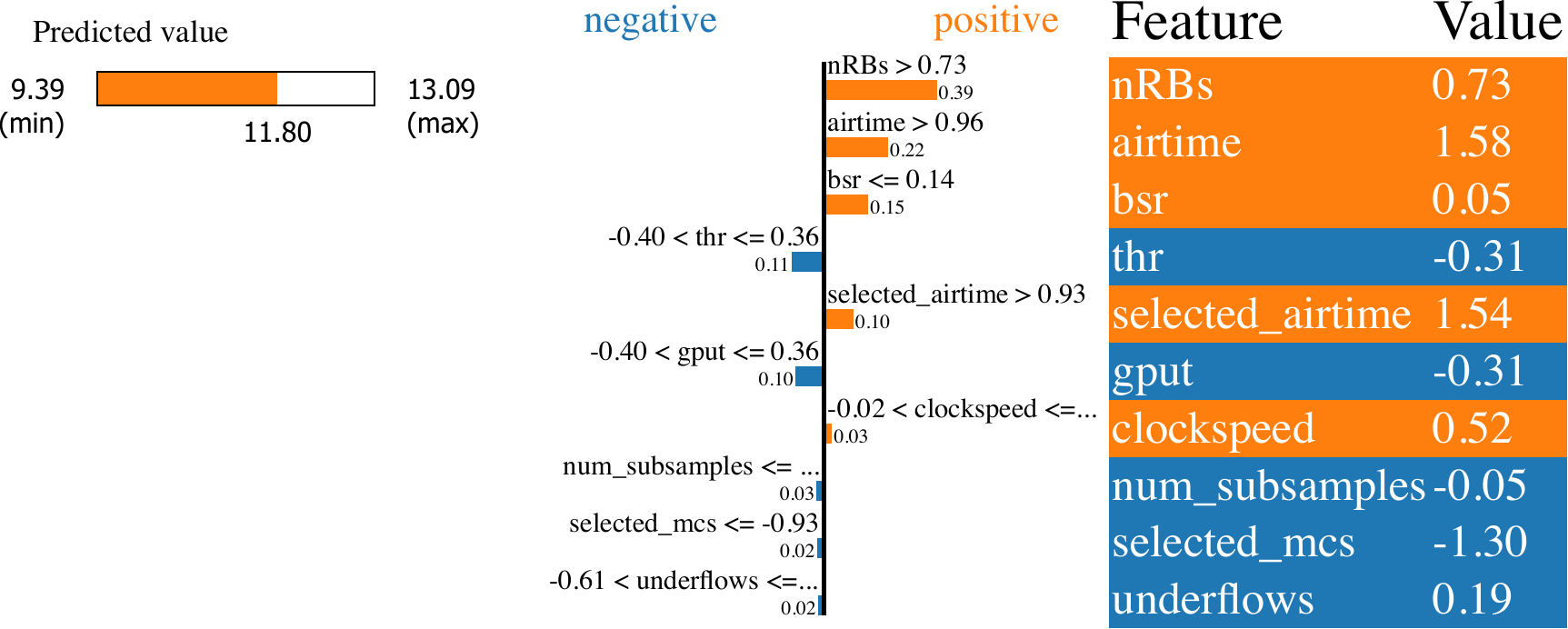}
       \caption{LIME: UL dataset using XGBoost}
        \label{fig:lime6}
    \end{subfigure}
    \begin{subfigure}{0.4965\linewidth}
        \centering
        \includegraphics[width=\linewidth]{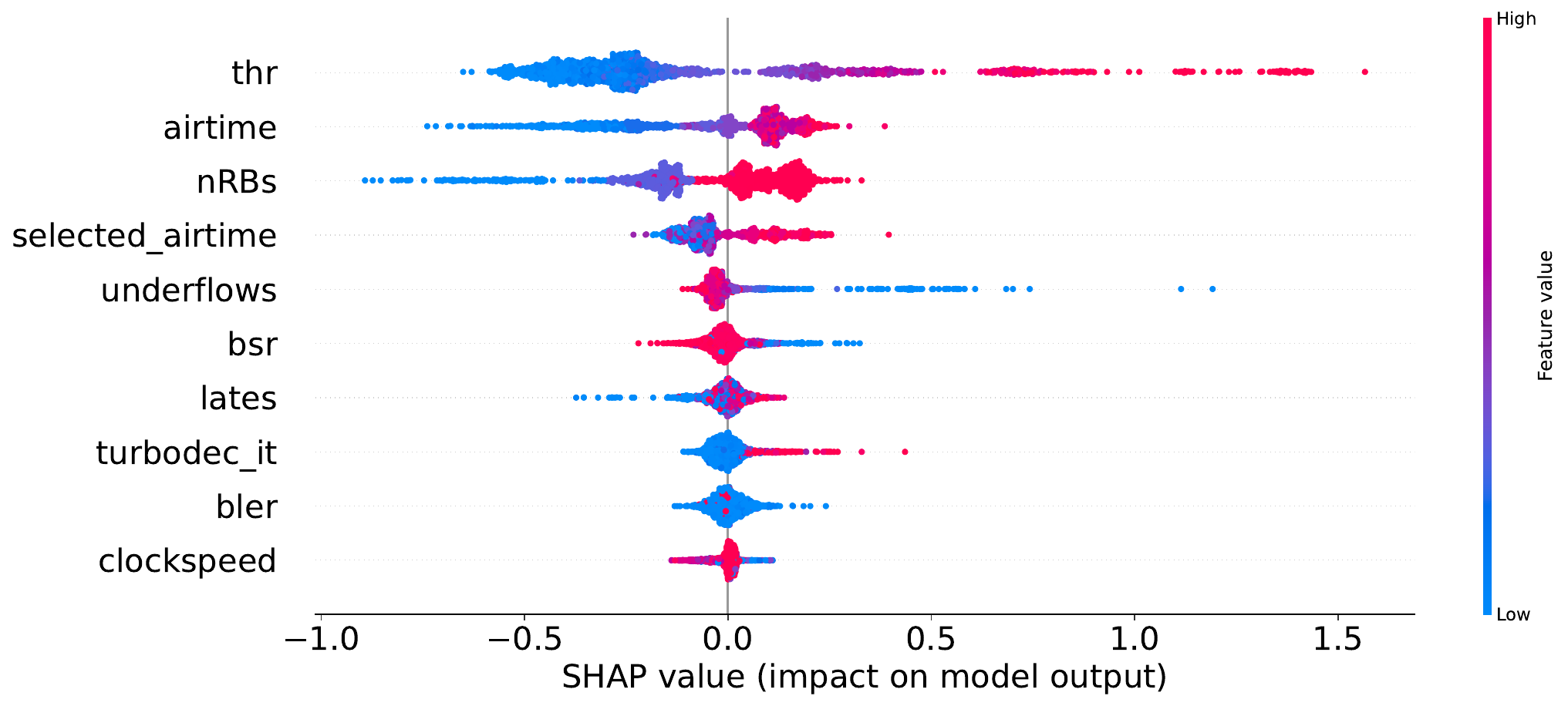}
        \caption{SHAP: UL dataset using XGBoost}
        \label{fig:shap6}
    \end{subfigure}
    \qquad
  \vspace{-0.3cm}    
    \caption{LIME and SHAP analysis for only UL RAN parameters influence on power consumption}
    \label{fig:shap}
\vspace{-0.4cm}
\end{figure*}

%\textcolor{blue}{
\begin{itemize}
    \item \emph{airtime} : The duration of time a base station is actively transmitting data over the air.
     \item \emph{number of Radio Blocks (nRB)} :  The quantity of radio resource units allocated for data transmission in a wireless network to determine the amount of bandwidth used per subframe, affecting network performance and resource utilization.
     \item \emph{Buffer Status Report (bsr)} : Amount of data waiting to be sent from a user's device to the base station.
     \item \emph{Goodput (gput)} :  Measures the amount of useful data successfully delivered to the destination per unit of time, excluding any retransmitted or corrupted packets.
     \item \emph{selected\_airtime} : The specific amount of time during which a wireless communication channel is actively used for transmitting data.
     \item \emph{throughput (thr)} : The rate at which data is successfully transmitted over a network, usually measured in bits per second (bps) --- Megabits per second (Mbps) or Gigabits per second (Gbps).
     \item \emph{Decoding time (dec\_time)}: The time required to decode data blocks in a wireless communication system.
     \item \emph{Block Error Rate (BLER)}: A measure of the reliability of data transmission, indicating the percentage of data blocks that are received with errors and require re-transmission. 
      \item \emph{Transmission Gain (tx\_again)}: The amplification level applied to a signal before it is transmitted by an antenna in a wireless communication system.
     \item \emph{turbodecoder iterations (turbodec\_it)}: The number of iterations performed by the turbo decoder during the error correction process.
\end{itemize}
%}
Fig.~\ref{fig:dlul} and Fig.~\ref{fig:shap} show the impact of various parameters of the \ac{RAN} and their power consumption for both DL/UL and UL datasets. 
The obtained influencing parameters between \ac{SHAP} and \ac{LIME} are different due to differences in their algorithms, approaches to handling feature interactions, and sensitivity to data perturbations and noise.
However, the considered models provide insights into the parameters influencing energy consumption.   

%%%%%%%%%%%%%%%%%%%%%%%%%%%%%%%%%%%%%%%%%%%%%%%%%%%%%%%%%%%%%%%%%%%%%%%%%%%

Both \ac{LIME} and \ac{SHAP} in all the considered AI/ML models report that (i) \emph{airtime}; %\textcolor{red}{(ii) \emph{number of Radio Blocks (nRBs)};} 
(ii) \emph{Average Buffer Status Report (bsr)}; (iii) \emph{Average Goodput (gput) uplink}; and (iv) \emph{selected\_airtime} are %top-5 
top-4 key parameters that could influence energy consumption of the \ac{RAN} for the DL/UL dataset as shown in Fig.~\ref{fig:dlul}. 

As shown in Fig.~\ref{fig:lime1}, the predicted value for the specific instance is reported in the top-left corner (13.53W).
The middle section shows how different features contribute to the model prediction by dividing it into negative and positive contributions.
The positive contribution of each feature --- the length of each feature bar represents the influence of feature contribution in energy consumption --- a longer bar means the feature exhibits a strong contribution to energy consumption.
The positive contributed features are represented in orange color and the negative contributed features are represented in blue color. 

The middle section values are ordered by their significant contributions. For example, longer \emph{airtime} %\textcolor{red}{and increased \emph{RBs} usage}
indicate higher network activity (i.e., active radio transceivers), leading to elevated energy consumption. 
The top right corner table shows the exact values of the considered instance.
% \textcolor{blue}{Airtime reflects the duration for which a user or service is
% actively using the network to transmit or receive data. Longer
% airtime suggests that radio transceivers are active for extended
% periods, consuming power to maintain data transmission.}.

%\textcolor{red}{Higher \emph{bsr} values suggest congestion, requiring more resources and energy. Improved \emph{gput} efficiency may reduce power consumption per unit of data. Optimal scheduling decisions represented by selected \emph{airtime} can also impact energy efficiency.} On the right, the actual feature values of the energy consumption are reported. 

Fig.~\ref{fig:shap1} shows the SHAP values plot generated from the Gradient Boosting model, the features \emph{dec\_time}, \emph{airtime\_ul}, \emph{thr\_ul}, \emph{gput\_ul}, \emph{thr\_dl}, and \emph{turbodec\_it} have a significant positive impact on energy consumption. This indicates that higher values of these features lead to increased energy consumption. In addition, \emph{selected\_airtime\_dl} and \emph{txgain\_ul} also contribute to energy consumption, to a lesser extent. However, \emph{bler\_ul} negatively impacts energy consumption, meaning that an increase in \emph{bler\_ul} results in a reduction in energy consumption.

Also, Fig.~\ref{fig:lime3} and \ref{fig:shap3} demonstrate the \ac{RAN} energy consumption for the DL/UL dataset using \emph{Random Forest}. 
The parameters such as \emph{airtime}, \emph{BSR}, \emph{thr}, \emph{gput}, 
%\textcolor{red}{and \emph{nRBs} }
significantly influence the \ac{RAN} power consumption.
Fig.~\ref{fig:lime3} shows longer \emph{airtime} and higher \emph{BSR} indicate increased network activity, demanding more energy.
Higher \emph{throughput} and \emph{Goodput} imply efficient data transfer, potentially lowering power consumption per data unit as they are on the negative side. 
 %\textcolor{red}{The number of RBs (nRBs) used correlates with network workload, which impacts power consumption.}

Fig.~\ref{fig:shap3} shows the SHAP values plot generated from the \emph{Random Forest} model --- \emph{thr\_ul}, \emph{gput\_ul}, \emph{turbodec\_it}, \emph{airtime\_ul}, \emph{thr\_dl}, and \emph{airtime\_dl} --- have a significant impact on energy consumption, indicating that higher values of these features lead to increased energy usage. The \emph{selected\_airtime\_ul} and \emph{selected\_airtime\_dl} have a smaller impact on energy consumption. Initially, \emph{bler\_dl} negatively impacts energy consumption, but as \emph{bler\_dl} increases, it starts to positively impact energy consumption. On the other hand, \emph{bsr\_ul} consistently has a negative impact on energy consumption. The intensity on the graph indicates regions where more data points are concentrated.

Fig.~\ref{fig:lime5} shows the LIME values plot generated from the \emph{XGBoost} model.
Here, the \emph{airtime\_ul} has a significant impact on energy consumption.
In contrast, \emph{selected\_airtime\_ul}, %\textcolor{red}{\emph{cqi\_ul}},
\emph{turbodec\_it}, \emph{bsr\_ul}, and \emph{selected\_mcs\_dl} have minimal impact on consumption.
All other parameters have a negative impact on energy consumption. 
Whereas, Fig.~\ref{fig:shap5} shows the SHAP values plot generated from the XGBoost model, the parameters --- \emph{gput\_ul}, \emph{dec\_time}, \emph{airtime\_ul}, \emph{turbodec\_it}, \emph{thr\_dl}, \emph{thr\_ul}, \emph{airtime\_dl}, and \emph{selected\_airtime\_ul} --- exhibit a significant impact on energy consumption, indicating that higher values of these features lead to increased energy usage. 
However, \emph{selected\_airtime\_dl} shows minimal impact on energy consumption.
Understanding how different features affect energy consumption helps in designing energy-efficient network systems.
For example, here airtime ul has a significant positive impact on power consumption which can urge to design protocols that minimize unnecessary airtime usage.

Fig.~\ref{fig:lime2} and~\ref{fig:shap2} depict the \ac{RAN} key parameters for the UL data using \emph{Gradient Boosting}. As similarly observed in Fig.~\ref{fig:dlul}, the UL-only data also shows \emph{airtime}, \emph{selected\_airtime}, \emph{nRBs} has a significant impact on energy consumption. In contrast, \emph{gput}, and \emph{num\_subsamples} have minimal impact on consumption. Most of all other parameters negatively affect energy consumption. Fig.~\ref{fig:shap2} shows the SHAP values plot generated from the \emph{Gradient Boosting} model, \emph{thr}, \emph{gput}, \emph{turbodec\_it}, \emph{airtime}, \emph{nRBs}, and \emph{selected\_airtime} show a significant impact on energy consumption. Initially, \emph{bler} has a negative impact, but as it increases, its impact turns positive. \emph{Underflows}, \emph{bsr}, and \emph{lates} consistently have a negative impact on energy consumption.

%%%%%%%%%%%%%%%RANDOM FOREST(UL)%%%%%%%%%%%%%%%%%%%%%%%
Fig.~\ref{fig:lime4} and~\ref{fig:shap4} depict the \ac{RAN} key parameters for the UL data using \emph{Random Forest}. Fig.~\ref{fig:lime4} shows the LIME values plot generated from the Random Forest model, \emph{nRBs}, \emph{airtime}, \emph{selected\_airtime}, and \emph{bsr} positively impact energy consumption, while all other parameters negatively affect energy consumption. Fig.~\ref{fig:shap4} shows the SHAP values plot generated from the Random Forest model, \emph{thr}, \emph{gput}, \emph{nRBs}, \emph{turbodec\_it}, \emph{selected\_airtime}, and \emph{airtime} exhibit a significant impact on energy consumption, whereas \emph{clockspeed} shows minimal impact. 
%Additionally, \emph{bsr}, \emph{lates}, and \emph{underflows} consistently have a negative impact on energy consumption.
%%%%%%%%%%%%%%%XGBOOST(UL)%%%%%%%%%%%%%%%%%%%%%%%

Finally, Fig.~\ref{fig:lime6} and~\ref{fig:shap6} depict the \ac{RAN} key parameters for the UL data using \emph{XGBoost model}.
Fig.~\ref{fig:lime6} shows the LIME values --- \emph{nRBs}, \emph{airtime}, \emph{bsr}, and \emph{selected\_airtime} --- have a positive impact on energy consumption, whereas \emph{clock speed} has minimal impact.
Fig.~\ref{fig:shap6} shows the SHAP values plot generated from the XGBoost model, \emph{thr} has a very high impact on energy consumption.
Following \emph{thr}, \emph{turbodec\_it}, \emph{selected\_airtime}, \emph{nRBs}, and \emph{airtime}
demonstrate a high impact on energy consumption.
Conversely, \emph{underflows}, \emph{bsr}, \emph{bler}, and \emph{clockspeed} all show a negative impact on energy consumption.

In summary, the analysis of \ac{RAN} energy consumption highlights key parameters --- \emph{airtime},
%\textcolor{red}{ \emph{Radio Blocks}},
\emph{BSR}, \emph{Goodput}, \emph{selected airtime} ---- %\textcolor{red}{\emph{CQI}} 
as significant influencers.
Longer \emph{airtime} 
%\textcolor{red}{and increased \emph{Radio Blocks} usage }
indicate higher network demand and energy usage, while higher \emph{BSR} values suggest congestion.
Conversely, improvements in \emph{Goodput} efficiency can reduce energy per unit of data.
%\textcolor{red}{Optimal scheduling decisions represented by selected airtime also impact energy efficiency. Understanding these parameters informs strategies to optimize energy usage and enhance network performance in \ac{RAN} deployments.}
This work serves as a benchmark for further research into reducing the carbon footprint of networks.
By establishing a clear connection between network parameters and power consumption, it sets the stage for developing more energy-efficient network architectures and protocols.

\section{conclusions and future work}
\vspace{-0.05cm}
This study explores various eXplainable AI (XAI) techniques such as LIME and SHAP and analyzes the impact of different RAN parameters on energy consumption. 
The considered XAI techniques are evaluated within a real-time RAN dataset and reports that variations in RAN parameters --- \emph{airtime}, \emph{goodput}, \emph{throughput}, \emph{subframe decoding time}, \emph{buffer status report}, \emph{number of resource blocks} --- could impact the RAN energy consumption.
%Also, the study explores the XAI integration for XR use case.
In future work, efforts may focus on refining XAI techniques, integrating them into O-RAN testbed for XR use case, and conducting further experiments to validate their effectiveness in optimizing energy usage and improving network performance as r/xApps.

\section*{Acknowledgments}
\vspace{-0.05cm}
This work has been partially supported by TTDF “SMART-RIC6G: Smart Drift-Handling Enabler for RAN Intelligent Controllers in 6G Networks (TTDF/6G/422)” project. 

%\section*{Acknowledgment}
%\vspace{-0.15cm}
%This work received funding from DST SERB Startup Research Grant (SRG-2021-001522), the SGNF project (``Reliability Evaluation of Virtualised 5G'').

\bibliographystyle{IEEEtran}
\vspace{-0.05cm}
\bibliography{biblio}
\end{document}